\documentclass[11pt]{article}

\usepackage{amsmath,amssymb,amsthm,amscd,latexsym}
\usepackage{hyperref}
\usepackage[toc,title,titletoc]{appendix}
\usepackage{cite}
\usepackage{graphicx,epsfig}
\RequirePackage{color}

\allowdisplaybreaks[1]

\def\nk{n_{\rm b}}
\def\sq{\sqrt{1-\ee}}
\def\wq{\ton{\wx^2 - \wy^2}}

\def\cp{\ton{3 + \cII}}
\def\wl{\ton{\wx^2 + \wy^2 - 2 \wz^2}}
\def\wpl{\ton{\wx\cO + \wy\sO}}
\def\wmn{\ton{\wy\cO - \wx\sO}}
\def\ple{\ton{- 2 + \ee + 2 \sq}}
\def\kle{\ton{- 1 + \ee +  \sq}}

\def\rfr#1{eq. (\ref{#1})}

\def\derp#1#2{\rp{\partial{#1}}{\partial{#2}}}
\def\dert#1#2{\frac{{{d}}{#1}}{{{d}}{#2}}}

\def\virg#1{``#1''}

\def\eqi{\begin{equation}}
\def\eqf{\end{equation}}
\def\eqia{\begin{eqnarray}}
\def\eqfa{\end{eqnarray}}
\def\Om{\mathit{\Omega}}
\def\rp#1#2{{#1\over#2}}
\def\lb#1{\label{#1}}

\def\wx{\hat{w}_x}
\def\wy{\hat{w}_y}
\def\wz{\hat{w}_z}
\def\bds#1{\boldsymbol{#1}}

\def\co{\cos\omega}
\def\so{\sin\omega}
\def\coo{\cos 2\omega}
\def\soo{\sin 2\omega}
\def\cO{\cos\Om}
\def\sO{\sin\Om}
\def\cOO{\cos 2\Om}
\def\sOO{\sin 2\Om}

\def\cI{\cos I}
\def\sI{\sin I}
\def\cII{\cos 2I}
\def\sII{\sin 2I}

\def\ee{e^2}

\def\ton#1{\left(#1\right)}
\def\qua#1{\left[#1\right]}
\def\grf#1{\left\{#1\right\}}
\def\ang#1{\left\langle #1\right\rangle}

\begin{document}

\title{Constraining the  Preferred-Frame $\alpha_1,\alpha_2$ parameters from  Solar System planetary precessions}

\author{L. Iorio\\ Ministero dell'Istruzione, dell'Universit$\grave{\textrm{a}}$ e della Ricerca (M.I.U.R.)-Istruzione \\ Fellow of the Royal Astronomical Society (F.R.A.S.)\\ Viale Unit$\grave{\textrm{a}}$ di Italia 68, 70125, Bari (BA), Italy}

\maketitle

\begin{abstract}
Analytical expressions for the orbital precessions affecting the relative motion of the components of a local binary system induced by Lorentz-violating Preferred Frame Effects (PFE)  are explicitly computed in terms of the PPN parameters  $\alpha_1,\alpha_2$.
\textcolor{black}{Preliminary} constraints on $\alpha_1,\alpha_2$ are inferred from \textcolor{black}{the} latest determinations of \textcolor{black}{the observationally admitted ranges $\Delta\dot\varpi$ for any anomalous Solar System planetary perihelion precessions}.
Other bounds existing in the literature are critically reviewed, with particular emphasis on the constraint $|\alpha_2|\lessapprox 10^{-7}$  based on an interpretation of the current close alignment of the Sun' s equator with the invariable plane of the Solar System in terms of the action of a $\alpha_2$-induced torque throughout the entire Solar System's existence.
Taken individually, the supplementary precessions $\Delta\dot\varpi$ of  Earth and Mercury, recently determined with the INPOP10a ephemerides \textcolor{black}{without modeling PFE}, yield  $\alpha_1 = (0.8\pm 4)\times 10^{-6}$ and $\alpha_2 = (4\pm 6)\times 10^{-6}$, respectively. A linear combination of the supplementary perihelion precessions of all the inner planets of the Solar System, able to remove the a-priori bias of unmodelled/mismodelled standard effects such as the general relativistic Lense-Thirring precessions and the classical rates due to the Sun's oblateness $J_2$, allows to infer $\alpha_1 = (-1\pm 6)\times 10^{-6},\alpha_2 = (-0.9\pm 3.5)\times 10^{-5}$. \textcolor{black}{Such figures are obtained by assuming that the ranges of values for the anomalous perihelion precessions are entirely due to the unmodeled effects of $\alpha_1$ and $\alpha_2$}. Our bounds should be improved in the near-mid future with the  MESSENGER and, especially, BepiColombo spacecrafts. \textcolor{black}{Nonetheless, it is worthwhile noticing that our constraints are close to those predicted for BepiColombo in two independent studies}. In further dedicated \textcolor{black}{planetary} analyses, PFE may be explicitly modeled to  estimate $\alpha_1,\alpha_2$ \textcolor{black}{simultaneously with the other PPN parameters as well}.
\end{abstract}

\centerline




\section{Introduction}
The invariance of the currently accepted laws of physics under Lorentz transformations of the spacetime degrees of freedom \cite{Kib61} is one of the most far-reaching ingredients of our vision of the physical reality. To date, it is routinely corroborated with the greatest accuracy in the high-energy realms of particle physics \cite{2011RvMP...83...11K}.  The current level of confidence we have on its validity in gravitational physics \cite{1993tegp.book.....W} is relatively  less satisfactory because of the challenges in accurate experiments involving gravity \cite{2001LRR.....4....4W}.

Modified models of the gravitational interaction encompassing violations of the Lorentz symmetry arise in several theoretical scenarios \cite{2005LRR.....8....5M,Capoz011,2012PhR...513....1C} such as, e.g.,  vector-metric theories \cite{1972ApJ...177..757W}, Tensor-Vector-Scalar (TeVeS) theories \cite{2004PhRvD..70h3509B}, Einstein-{\AE}ther theories \cite{2001PhRvD..64b4028J,2005PhRvD..72b5012H,Jac07,2007PhRvD..75d4017Z,2012CoTPh..57..227M}, MOND \cite{2006ConPh..47..387B,2007PhRvD..76l4012B,2011PhRvD..84h4024S,2012LRR....15...10F}, Ho{\v r}ava-Lifshitz-type theories \cite{2009PhRvD..79h4008H,2009JHEP...03..020H,2010PhRvD..81j1502J,2011JPhCS.283a2034S,2012PhRvD..85j5001P}, supersymmetric field theories \cite{2005PhRvL..94h1601N,2012JHEP...01..062P}, Standard Model Extensions (SME) \cite{2004PhRvD..69j5009K,2006PhRvD..74d5001B,2006PhLB..642....9L,2010PhRvD..82f5012B,2011PhRvD..83a6013K}.
They predicts several Preferred Frame Effects (PFE)
due to the existence of a putative Preferred Frame (PF) which might be singled out by, e.g., the Universe matter distribution.

In the framework of the Parameterized Post-Newtonian (PPN) formalism, Preferred Frame Effects (PFE) are phenomenologically taken into account by the PPN parameters $\alpha_1,\alpha_2,\alpha_3$ \cite{1972ApJ...177..757W}. Since $\alpha_3$ is also related to possible violations of the matter-energy conservation and is very accurately constrained down to a $\approx 10^{-20}$ level from pulsar acceleration data \cite{2005ApJ...632.1060S}, it will be neglected in the following. However, caution is required in straightforward extension of strong-field constraints to  weak-field scenarios: $\alpha_3$ will be the subject of a forthcoming dedicated paper. For an overall overview on the present-day constraints on all the PPN parameters, see \cite{2001LRR.....4....4W}. As far as $\alpha_1,\alpha_2$ are concerned, see also the discussion in Section \ref{osserva}; at present, they are constrained at $\approx 10^{-4}\ (\alpha_1),\approx 10^{-7}\ (\alpha_2)$ level by a variety of approaches.

In this paper, we intend to \textcolor{black}{preliminarily explore  a different approach with respect to those usually followed so far\footnote{\textcolor{black}{See, e.g., \cite{2001LRR.....4....4W} and references therein.}} in constraining both $\alpha_1$ and $\alpha_2$}.

The plan of the work is as follows. In Section \ref{hamiltoniane} we work out the $\alpha_1,\alpha_2$  Hamiltonians. They are the basis for perturbatively calculating the averaged orbital precessions of the relative motion of the binary's components in Section \ref{rates} by using the Lagrange planetary equations. The resulting analytical formulas are exact in the sense that no a-priori simplifying assumptions about the PF velocity and the orbital geometry are made. In Section \ref{osserva}, after critically discussing the constraints on $\alpha_1,\alpha_2$ existing in literature, we use the latest determinations of the orbital motions of some planets of the Solar System to \textcolor{black}{preliminary} infer  bounds on both $\alpha_1$ and $\alpha_2$. \textcolor{black}{More specifically, we compare our analytical predictions with the latest ranges of values for any conceivable anomalous perihelion precessions; since the latter ones are statistically compatible with zero, we are able to obtain upper bounds on $\alpha_1,\alpha_2$. In doing that, we assume that the supplementary precessions are entirely due to PFE, not modelled in the dynamical theories fitted by the astronomers to the data. In other words, we actually test PFE theories departing from general relativity in at most $\alpha_1,\alpha_2$. Nonetheless, it turns out that our constraints are close to those independently predicted for the future BepiColombo mission by simultaneously estimating $\alpha_1,\alpha_2$ along with other PPN parameters from simulated data \cite{2002PhRvD..66h2001M,2007PhRvD..75b2001A}.} Section \ref{concludi} summarizes our findings.
\section{The perturbing Hamiltonians}\lb{hamiltoniane}
Let us consider a local binary system of total mass $M=m_{\rm A}+m_{\rm B}$.
Let $\bds r^0_{\rm CM} $ be the position vector of the system's center of mass (CM) with respect to the origin of the PF.
Thus, we can write
\begin{align}
\bds r^0_{\rm A} & = \bds r^{'}_{\rm A} + \bds r^0_{\rm CM} = -\rp{m_{\rm B}}{M}\bds r + \bds r^0_{\rm CM}, \\ \nonumber \\
\bds r^0_{\rm B} & = \bds r^{'}_{\rm B} + \bds r^0_{\rm CM}  = \rp{m_{\rm A}}{M}\bds r + \bds r^0_{\rm CM}
\end{align}
where the index $'$ refers to the CM frame; the position vector of $m_{\rm B}$ with respect to $m_{\rm A}$ is
$\bds r = \bds r_{\rm B}^0-\bds r_{\rm A}^0$  and $\bds{\hat{r}}=\bds r /r$ is its unit vector; \textcolor{black}{in the following, a  hat symbol  will always denote unit vectors.} It follows
\begin{align}
\bds v^0_{\rm A} \lb{velosA} &= -\rp{m_{\rm B}}{M}\bds v + \bds w, \\ \nonumber \\
\bds v^0_{\rm B} \lb{velosB} &= \rp{m_{\rm A}}{M}\bds v + \bds w,
\end{align}
where $\bds w$ is the  PF velocity of the CM.

At post-Newtonian level, the PFE two-body reduced\footnote{It is the ratio of the Lagrangian to the reduced mass $\mu \doteq m_{\rm A}m_{\rm B}/M$ of the binary system.} Lagrangian consists of the sum of the following terms \cite{1985ApJ...297..390N,1987ApJ...320..871N,1992PhRvD..46.4128D,2012arXiv1209.4503S}
\begin{align}\lb{lags1}
\mathcal{L}_{\alpha_1} & = -\rp{\alpha_1 G M }{2 c^2 r}\ton{\bds v^0_{\rm A}\bds\cdot\bds v^0_{\rm B}}, \\ \nonumber \\
\mathcal{L}_{\alpha_2} & = -\rp{\alpha_2}{\alpha_1}\mathcal{L}_{\alpha_1} +\widetilde{{\mathcal{L}}}_{\alpha_2},\lb{lags2}
\end{align}
with
\eqi \widetilde{{\mathcal{L}}}_{\alpha_2}\doteq -\rp{\alpha_2 G M}{2c^2 r}\ton{\bds v^0_{\rm A}\bds\cdot\bds{\hat{r}}}\ton{\bds v^0_{\rm B}\bds\cdot\bds{\hat{r}}},\lb{lagga}
\eqf
where $G$ is the Newtonian constant of gravitation, and $c$ is the speed of light in vacuum. For a comparison with the bounds on $\alpha_2$ inferred by Nordtvedt \cite{1985ApJ...297..390N,1987ApJ...320..871N}, we notice that he adopted a nonstandard normalization: $\alpha_2^{\rm Nord} = \alpha_2 /2$.

Since we are interested in the relative orbital motion of the binary's components, we have to express \rfr{lags1}-\rfr{lagga} in terms of relative quantities\footnote{\textcolor{black}{The motion of the Sun with respect to the Solar System Barycenter (SSB) has been recently analyzed in \cite{2013arXiv1306.5569V}.}}.
From \rfr{velosA}-\rfr{velosB} it turns out,
\begin{align}\lb{relav1}
\bds v^0_{\rm A}\bds\cdot\bds v^0_{\rm B} & = w^2 + \rp{\Delta m}{M}\ton{\bds v\bds\cdot\bds w}-\rp{m_{\rm A} m_{\rm B}}{M^2}v^2, \\ \nonumber \\
\ton{\bds v_{\rm A}^0\bds\cdot\bds{\hat{r}}}\ton{\bds v_{\rm B}^0\bds\cdot\bds{\hat{r}}} &= w_r^2 + \rp{\Delta m}{M}w_r v_r - \rp{m_{\rm A} m_{\rm B}}{M^2}v_r^2,\lb{relav2}
\end{align}
where $v_r \doteq \bds v\bds\cdot\bds{\hat{r}}, w_r \doteq \bds w\bds\cdot\bds{\hat{r}}, \Delta m \doteq m_{\rm A} - m_{\rm B}$.
Thus, \rfr{lags1}-\rfr{lags2} become
\begin{align}\lb{ru1}
\mathcal{L}_{\alpha_1} & = -\rp{\alpha_1G M}{2 c^2 r}\qua{w^2 + \rp{\Delta m}{M}\ton{\bds v\bds\cdot\bds w}-\rp{m_{\rm A} m_{\rm B}}{M^2}v^2}, \\ \nonumber \\
\mathcal{L}_{\alpha_2} & = -\rp{\alpha_2}{\alpha_1}\mathcal{L}_{\alpha_1} + \widetilde{{\mathcal{L}}}_{\alpha_2},\lb{ru2}
\end{align}
with
\eqi\widetilde{{\mathcal{L}}}_{\alpha_2} = -\rp{\alpha_2 G M}{2c^2 r}\ton{w_r^2 + \rp{\Delta m}{M}w_r v_r - \rp{m_{\rm A} m_{\rm B}}{M^2}v_r^2}.\lb{lagtilde}\eqf
By recalling that a general velocity-dependent perturbative Lagrangian ${\mathcal{L}}_{\rm pert}$ yields
a corresponding perturbative Hamiltonian \cite{2004A&A...415.1187E,2005CeMDA..91...75E,2011rcms.book.....K}
\eqi {\mathcal{H}}_{\rm pert} = -{\mathcal{L}}_{\rm pert} - \rp{1}{2}\ton{\derp{{\mathcal{L}}_{\rm pert}}{\bds v}}^2, \eqf
the following PFE reduced Hamiltonians  can be obtained, to order $\mathcal{O}\ton{G/c^2}$, from \rfr{ru1}-\rfr{lagtilde}
\begin{align}\lb{ham1}
\mathcal{H}_{\alpha_1} &= \rp{\alpha_1 GM }{2 c^2 r}\qua{w^2  + \rp{\Delta m}{M}\ton{\bds v\bds\cdot\bds w} - \rp{m_{\rm A}m_{\rm B}}{M^2}v^2}, \\ \nonumber \\
\mathcal{H}_{\alpha_2} &= -\rp{\alpha_2}{\alpha_1}\mathcal{H}_{\alpha_1} + \widetilde{{\mathcal{H}}}_{\alpha_2}
\lb{ham2},
\end{align}
with
\eqi\widetilde{{\mathcal{H}}}_{\alpha_2} = \rp{\alpha_2 GM }{2c^2 r}\ton{w_r^2 + \rp{\Delta m}{M}v_r w_r - \rp{m_{\rm A}m_{\rm B}}{M^2}v_r^2}.\lb{hamtilde}\eqf
Perturbing accelerations can be obtained from \rfr{ham1}-\rfr{hamtilde} as
\eqi \dot{\bds p} = -\derp{{\mathcal{H}}}{{\bds r}}\lb{Hameq}\eqf by noting that
\begin{align}
\bds p_{\alpha_1} \lb{p1} & \doteq \derp{{\mathcal{L}_{\alpha_1}}}{{\bds v}} =-\rp{\alpha_1G M}{2 c^2 r}\ton{  \rp{\Delta m}{M}\bds w - 2\rp{m_{\rm A} m_{\rm B}}{M^2}\bds v}, \\ \nonumber \\
\widetilde{\bds p}_{\alpha_2} \lb{p2} & \doteq \derp{\widetilde{\mathcal{L}}_{\alpha_2}}{{\bds v}} = -\rp{\alpha_2 GM}{2 c^2 r}\ton{\rp{\Delta m}{M}w_r  - 2\rp{m_{\rm A} m_{\rm B}}{M^2}v_r}\bds{\hat{r}}.
\end{align}
\section{Averaged orbital perturbations}\lb{rates}
An effective  method to analytically calculate the orbital effects induced by a generic small correction
${\mathcal{H}}_{\rm pert}$ to the Newtonian potential $U_{\rm N}=-GM/r$ consists of evaluating ${\mathcal{H}}_{\rm pert}$ over the unperturbed Keplerian ellipse, assumed as reference trajectory,  and averaging it over one orbital period $P_{\rm b}$ of the test particle. Then, the Lagrange perturbation equations \cite{befa,Capde,2011rcms.book.....K} allow to straightforwardly calculate to first order the long-term rates of change of the orbital elements by means of partial derivatives of $\ang{{\mathcal{H}}_{\rm pert}}$ with respect to them. In principle, it would be possible to adopt a different reference trajectory as unperturbed orbit including also general relativity at the 1PN level and use the so-called post Newtonian Lagrange planetary equations \cite{1997PhRvD..56.4782C,1998CQGra..15.3121C}. Nonetheless, in the specific case of \rfr{ham1}-\rfr{hamtilde}, in addition to the first order precessions of order $\mathcal{O}\ton{c^{-2}}$ coming from the use of the Newtonian reference trajectory to  be computed below,
other \virg{mixed} 1PN-PF precessions  of higher order would arise specifying the influence of PFE on the 1PN post-Keplerian  orbital motion assumed as unperturbed. From the point of view of constraining $\alpha_1,\alpha_2$ from observations, they are practically negligible since their magnitude is quite smaller than the first order terms.

Before performing the calculation, a consideration about the orbital elements employed in it is in order. Strictly speaking, since the perturbing Hamiltonians of \rfr{ham1}-\rfr{hamtilde} depend also upon the velocity $\bds v$, inserting them into the usual Lagrange variation equations yields nonosculating orbital elements called contact elements \cite{1971CeMec...3..197B}, which are not tangent to the perturbed trajectory \cite{2004A&A...415.1187E,2005NYASA1065..346E}. If certain general conditions on the periodicity of motion are satisfied \cite{2011rcms.book.....K}, the averaging procedure reduces the difference between the contact and the osculating elements in such a way that residual differences occur only to higher order. See \cite{1994PhRvD..49.1693D} for a preliminary discussion of this issue in the present context. As it will be discussed in Section \ref{osserva}, it is just what occurs in the present case.

The semimajor axis $a$, which is dimensionally a length characterizing the size of the orbit, varies according to the Lagrange equation \cite{befa,2011rcms.book.....K}
\eqi\ang{\dert a t} = -\rp{2}{n_{\rm b}a}\derp{\ang{{\mathcal{H}}_{\rm pert}}}{\textcolor{black}{\mathcal{M}_0}}.\lb{pippa1}\eqf
 In it, $n_{\rm b} \doteq \sqrt{GM/a^3}$ is the Keplerian mean motion in terms of which the orbital period is expressed as $P_{\rm b}=2\pi/\nk$; its dimensions are those of a frequency. The angle \textcolor{black}{$\mathcal{M}_0$ is the mean anomaly at the epoch $t_0$ \cite{2011rcms.book.....K}.}

The Lagrange equation for the eccentricity $e$, which is a dimensionless numerical parameter determining the shape of the ellipse in such a way that\footnote{A circular orbit corresponds to $e = 0$.} $0\leq e < 1$, is \cite{befa,2011rcms.book.....K}
\eqi
\ang{\dert e t} = \rp{1}{\nk a^2}\ton{\rp{1 - \ee}{e}}\qua{\rp{1}{\sqrt{1-\ee}} \derp{ \ang{ {\mathcal{H}}_{\rm pert} } }{\omega} -
\derp{ \ang{ {\mathcal{H}}_{\rm pert} } }{{\textcolor{black}{\mathcal{M}_0}}} }.\lb{pippa2}\eqf In it, the argument of pericenter $\omega$ is an angle in the orbital plane counted from the line of the nodes (see below) to the point of closest approach. Although not yet explicitly determined for Solar System's planets \cite{2011CeMDA.111..363F}, the rate of change of $e$  is actually measured in some binary pulsar systems \cite{2011MNRAS.412.2763F,2012CQGra..29r4007F,2012arXiv1209.4503S}.

The inclination $I$ of the orbital plane to the reference $\{x,y\}$ plane\footnote{It is the plane of the sky in the case of a celestial binary system such as, e.g., a binary pulsar, while it is customarily the Earth's mean equator at the epoch J2000.0 if the planets of our Sun are considered.}
Its rate of change has not yet been explicitly determined by the astronomers for planets of the Solar System \cite{2011CeMDA.111..363F}. On the other hand, its precession is one of the observable quantities in binary pulsar studies through its connection with the  projected semimajor axis $x_{\rm p}\doteq a_{\rm p}\sI/c,$ where $a_{\rm p}$ is the barycentric semimajor axis of the pulsar. Indeed, its rate of change $\dot x_{\rm p}$ is often accessible to observation in several binaries \cite{2011MNRAS.412.2763F,2012CQGra..29r4007F,2012arXiv1209.4503S}.
The  Lagrange equation for $I$ is \cite{befa,2011rcms.book.....K}
\eqi\ang{\dert I t} = \rp{1}{n_{\rm b}a^2\sqrt{1-\ee}\sI}\qua{\derp{\ang{{\mathcal{H}}_{\rm pert}}} {\mathit{\Omega}} - \cI\derp{\ang{{\mathcal{H}}_{\rm pert}}} \omega}.\eqf In it, the longitude of the ascending node $\mathit{\Omega}$ is an angle in the $\{x,y\}$ reference plane counted from the $x$ reference direction to the line of the nodes which, in turn, is the intersection of the orbital plane with the $\{x,y\}$  plane itself.

The node is one of the orbital parameters used by the astronomers \cite{2011CeMDA.111..363F} to constrain putative non-standard dynamical effects in the Solar System. Sometimes, it is measurable also in specific binary pulsars \cite{Kop95,Kop96,Strat01}. The long-term precession of $\mathit{\Omega}$ is computed from the Lagrange perturbation equation \cite{befa,2011rcms.book.....K}
\eqi\ang{\dert{\mathit{\Omega}} t} = -\rp{1}{n_{\rm b}a^2\sqrt{1-\ee}\sI}\derp{\ang{{\mathcal{H}}_{\rm pert}}} I.\lb{dNd}\eqf

The longitude of pericenter $\varpi\doteq \mathit{\Omega} + \omega$, which is a \virg{broken} angle, is another parameter usually  adopted in both Solar System and binary pulsars studies to put limits on putative exotic forces \cite{2011CeMDA.111..363F}.  Its Lagrange perturbation equation is \cite{befa,2011rcms.book.....K}
\eqi\ang{\dert\varpi t} = -\rp{1}{n_{\rm b}a^2}\qua{\ton{\rp{\sqrt{1-\ee}}{e}}\derp{\ang{{\mathcal{H}}_{\rm pert}}} e + \rp{\tan\ton{\rp{I}{2}}}{\sqrt{1-\ee}}\derp{\ang{{\mathcal{H}}_{\rm pert}}} I }.\lb{lagvarpi}\eqf

In principle, also the mean anomaly $\mathcal{M}=\mathcal{M}_0 + \nk(t-t_0)$ \cite{2011rcms.book.....K} should be considered, but it will not be treated here since its precession is neither determined in Solar System analyses of planetary motions nor in binary pulsars. Indeed, its accuracy is necessarily limited by the uncertainty in the gravitational parameter $GM$ entering $\nk$, making the use of $\mathcal{M}$ usually less competitive than the other orbital elements.

Other analytical calculations of the orbital effects of non-vanishing $\alpha_1, \alpha_2$, made with different computational strategies and based on various levels of approximation, can be found in, e.g., \cite{1972ApJ...177..775N,1992PhRvD..46.4128D,1994PhRvD..49.1693D,2007MNRAS.380..455W,2012arXiv1209.4503S}.

Our analytical expressions for the $\alpha_1,\alpha_2$ long-term precessions are displayed in Appendix \ref{La1} and Appendix \ref{La2}. As fast variable of integration in averaging \rfr{ham1}-\rfr{hamtilde}, it turned out computationally more convenient to adopt the eccentric anomaly $E$. \textcolor{black}{Nonetheless, as a further check, we repeated  the calculation with the true anomaly $f$ as well: the same results were re-obtained.}

%

%
%
%
%
\section{Confrontation with the observations}\lb{osserva}
The analytical results of Appendix \ref{La1} and Appendix \ref{La2} were successfully confirmed by numerically integrating the equations of motion of a test particle including the  PF accelerations coming from \rfr{Hameq} with \rfr{ham1}-\rfr{hamtilde} and \rfr{p1}-\rfr{p2} over a given time interval, and plotting the temporal evolution of the numerically computed orbital elements with the standard formulas of the osculating Keplerian elements. In Figure \ref{figura1} and Figure \ref{figura2} the plots for Mercury over one century, obtained for $\alpha_1=\alpha_2=1$ just for illustrative purposes and with \rfr{wmap} for $\bds w$, are displayed.
\begin{figure*}
\centering
\begin{tabular}{cc}
\epsfig{file=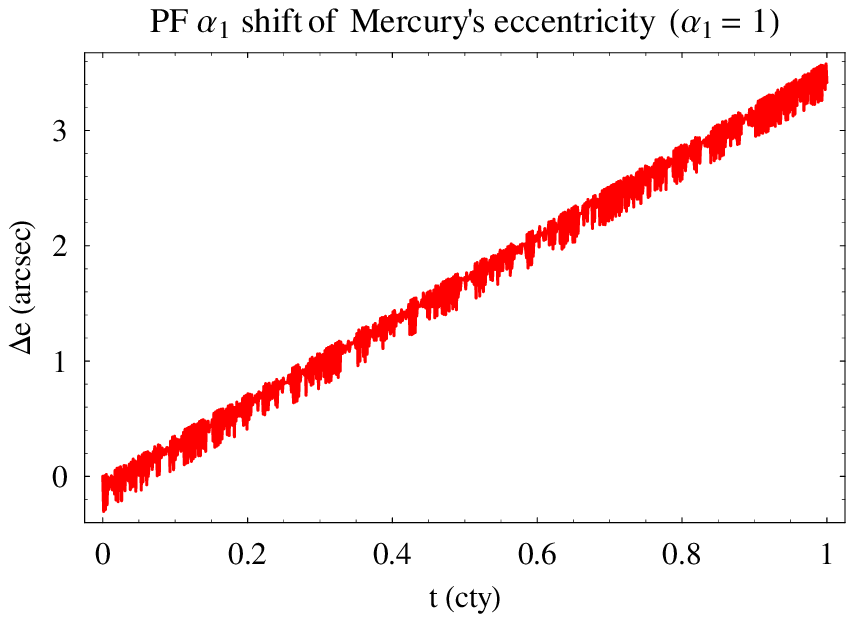,width=0.45\linewidth,clip=} & \epsfig{file=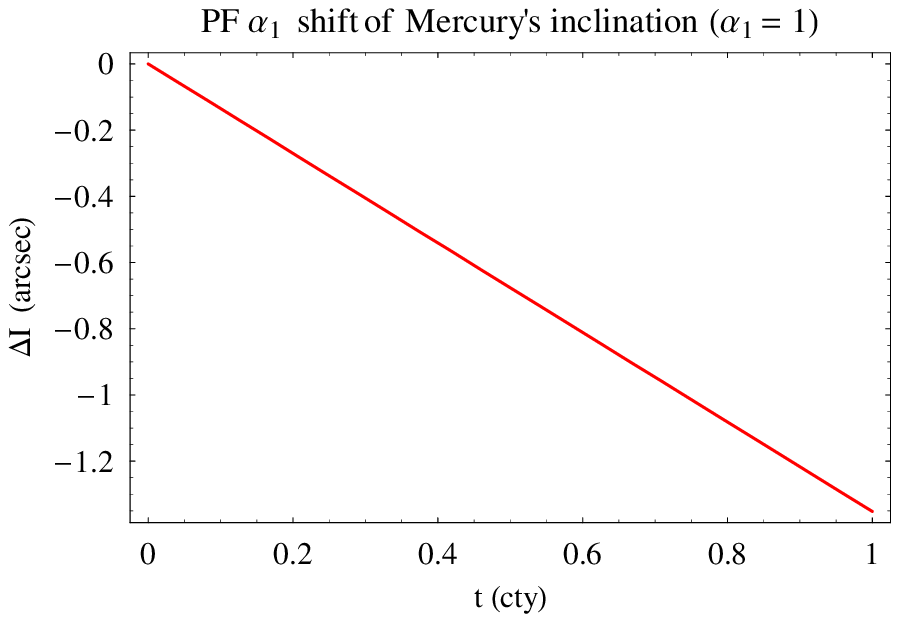,width=0.45\linewidth,clip=}\\
\epsfig{file=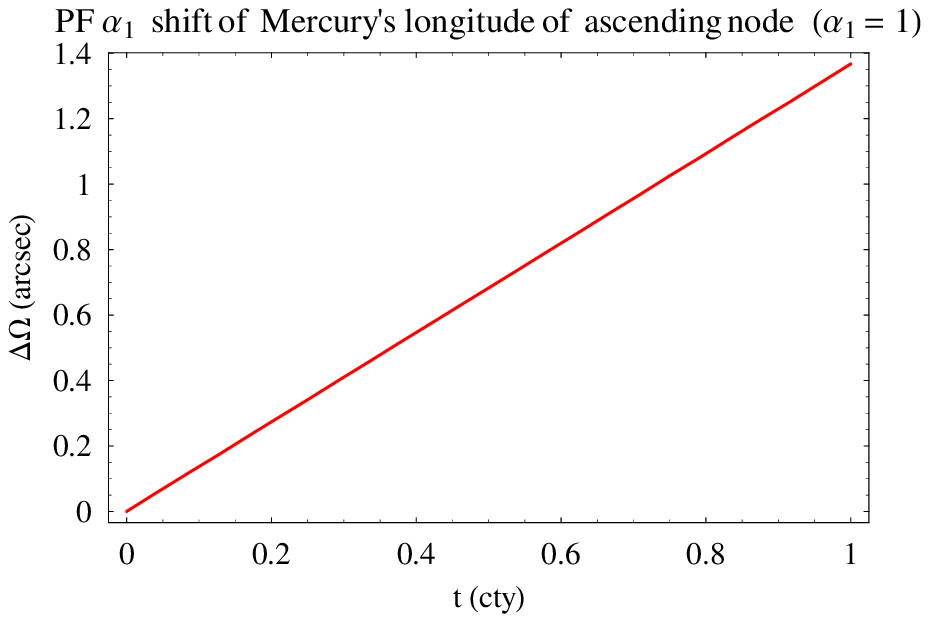,width=0.45\linewidth,clip=} & \epsfig{file=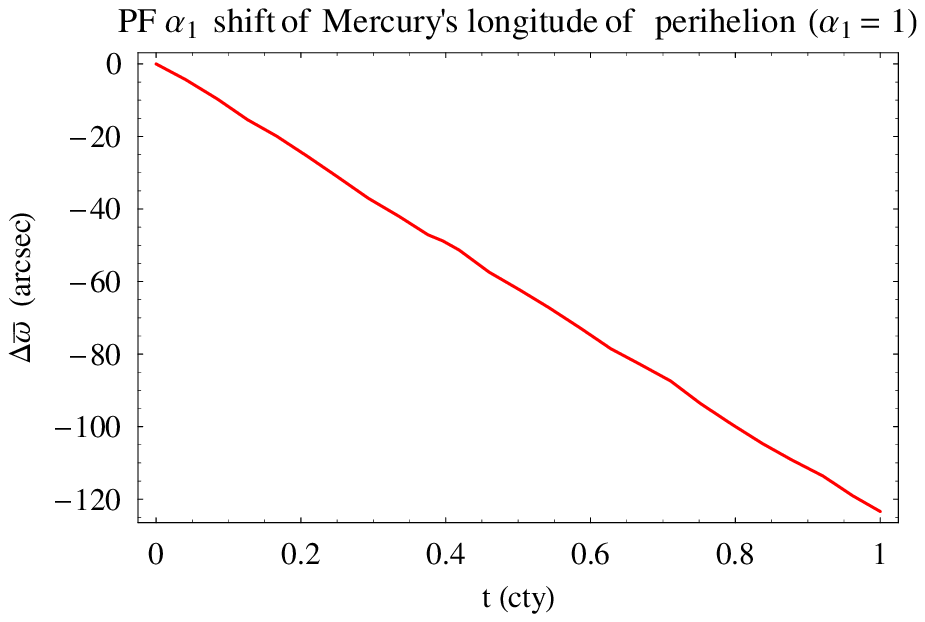,width=0.45\linewidth,clip=}\\
\end{tabular}
\caption{Numerically integrated temporal evolutions of the total PF $\alpha_1-$induced shifts (in arcsec) of the eccentricity $e$, the inclination $I$, the longitude of the ascending node $\mathit{\Omega}$ and the longitude of perihelion $\varpi$ of Mercury over one century. They were obtained by numerically integrating the planet's equations of motion in cartesian coordinates with and without the perturbative accelerations arising from \rfr{ham1} by using the same initial conditions retrieved from the NASA JPL HORIZONS System (http://ssd.jpl.nasa.gov/?horizons). For illustrative purposes the value $\alpha_1=1$ was chosen. The figures in \rfr{wmap}, taken from \cite{2009ApJS..180..225H}, were adopted for $\bds w$. The resulting centennial rates are in agreement with the sum of those computed from the analytical formulas in Appendix \ref{La1}.}\lb{figura1}
\end{figure*}
\begin{figure*}
\centering
\begin{tabular}{cc}
\epsfig{file=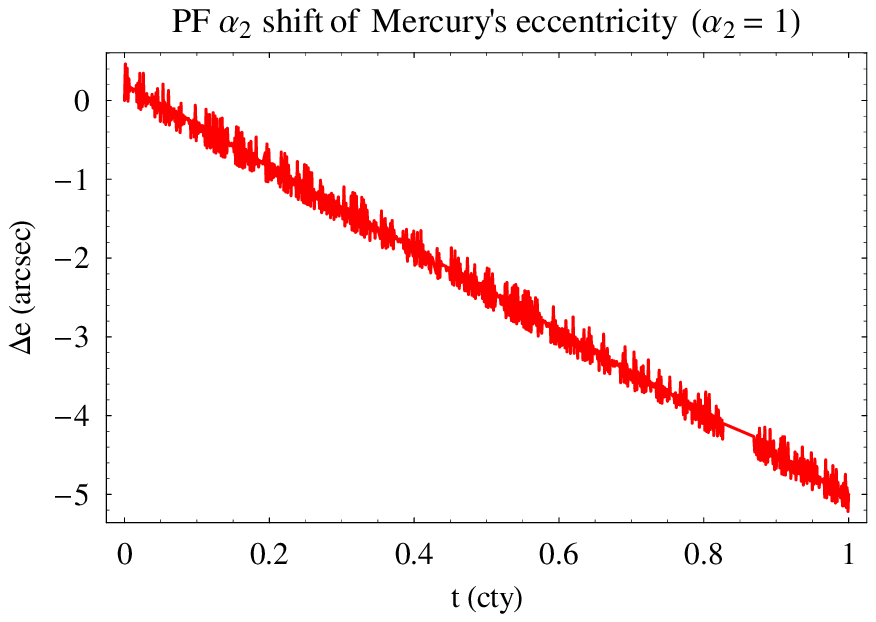,width=0.45\linewidth,clip=} & \epsfig{file=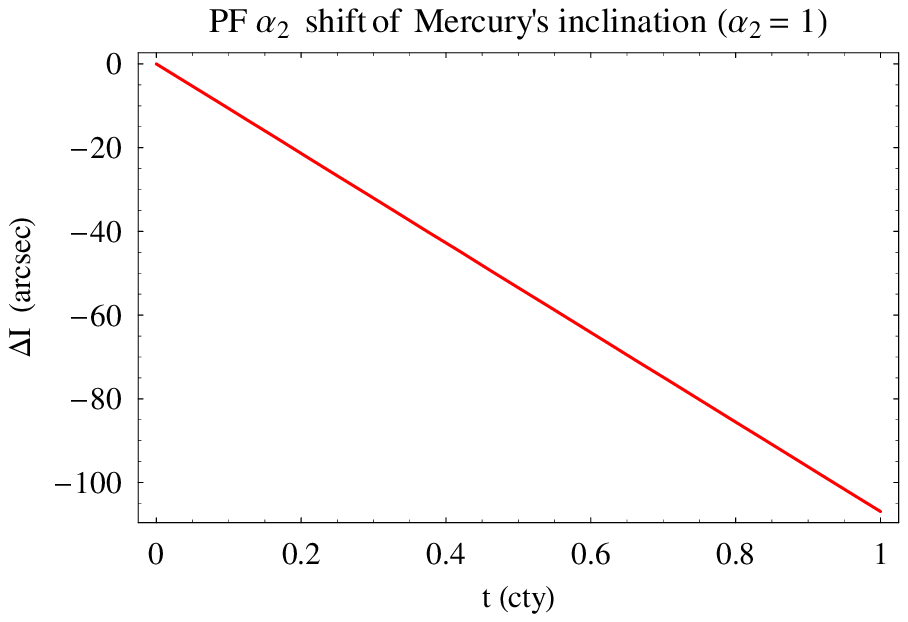,width=0.45\linewidth,clip=}\\
\epsfig{file=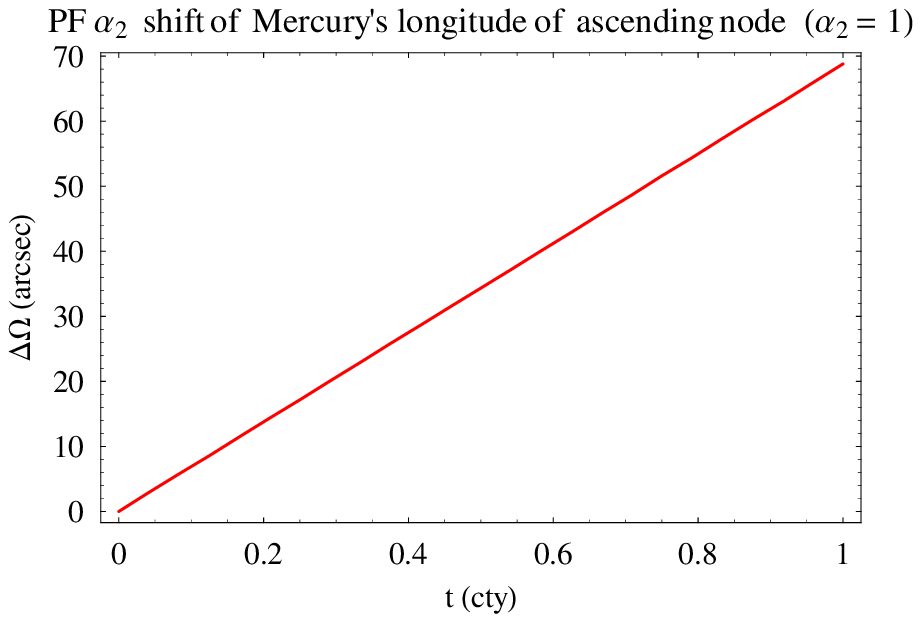,width=0.45\linewidth,clip=} & \epsfig{file=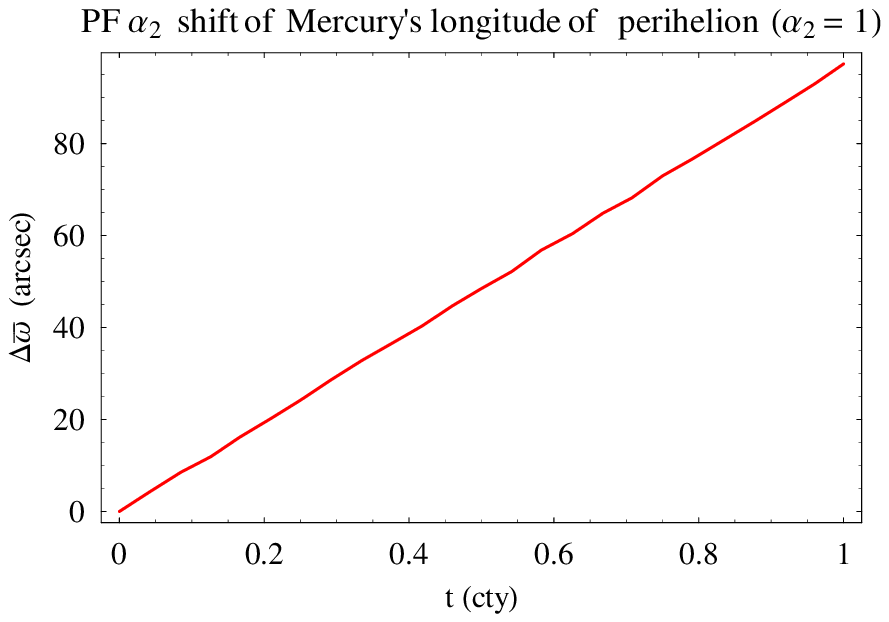,width=0.45\linewidth,clip=}\\
\end{tabular}
\caption{Numerically integrated temporal evolutions of the total PF $\alpha_2-$induced shifts (in arcsec) of the eccentricity $e$, the inclination $I$, the longitude of the ascending node $\mathit{\Omega}$ and the longitude of perihelion $\varpi$ of Mercury over one century. They were obtained by numerically integrating the planet's equations of motion in cartesian coordinates with and without the perturbative accelerations arising from \rfr{ham2}-\rfr{hamtilde} by using the same initial conditions retrieved from the NASA JPL HORIZONS System (http://ssd.jpl.nasa.gov/?horizons). For illustrative purposes the value $\alpha_2=1$ was chosen. The figures in \rfr{wmap}, taken from \cite{2009ApJS..180..225H}, were adopted for $\bds w$. The resulting centennial rates are in agreement with the sum of those computed from the analytical formulas in Appendix \ref{La2}.}\lb{figura2}
\end{figure*}
They yield the same centennial rates obtainable from the analytical formulas of Appendix \ref{La1} and Appendix \ref{La2} for the contact elements, computed for the same values of $\alpha_1,\alpha_2,\bds w$ used in the numerical integration. This shows that, to the approximation level used in the calculation, the contact elements are equal to the osculating elements.


Thus, confident in our results of Appendix \ref{La1}-Appendix \ref{La2}, we are going to apply them to effectively constrain both $\alpha_1$ and $\alpha_2$ from latest determinations of planetary orbital motions \cite{2011CeMDA.111..363F}. In Section \ref{liter} we critically review the bounds existing in literature.
\subsection{Discussion of the constraints existing in literature}\lb{liter}
As a general rule pertaining the bounds on  the strong-field equivalent $\hat{\alpha}_1,\hat{\alpha}_2$ with compact object, Shao and Wex \cite{2012arXiv1209.4503S} interestingly remark that their general validity  may be questioned by a potential compactness-dependence (or mass-dependence) of $\hat{\alpha}_1,\hat{\alpha}_2$ because of certain peculiar phenomena, such as spontaneous scalarization \cite{1993PhRvL..70.2220D}, which may take place. Thus, Shao and Wex \cite{2012arXiv1209.4503S} warn that it is always recommendable to specify the binary system used to infer given constraints on $\hat{\alpha}_1,\hat{\alpha}_2$. Another limitation of the strong-field tests is their probabilistic nature since the orbital configurations of the binaries used are not always completely known: indeed, the longitudes of the ascending nodes $\mathit{\Omega}$ are often unknown, and also the inclinations $I$ may be sometimes determined modulo the ambiguity of $I\rightarrow 180^{\circ} - I$.
\subsubsection{$\alpha_1$}
Earlier  constraints on $\alpha_1$ were obtained from Solar System planetary motions \cite{1972ApJ...177..775N,1984grg..conf..365H} and Lunar Laser Ranging \cite{LLR08}; the most accurate bounds \cite{1984grg..conf..365H,LLR08} were at $\approx 2\times 10^{-4}$ level.

Strong-field tests were reported by using compact objects such as neutron stars and white dwarfs \cite{1992PhRvD..46.4128D,2000ASPC..202..113W}. They were able to constrain $\hat{\alpha}_1$, which, in principle, may differ from the weak-field value $\alpha_1$ because of possible contributions from strong-field effects, down to $5.0-1.2\times 10^{-4}$. Latest bounds on $\hat{\alpha}_1$ were obtained by Shao and Wex \cite{2012arXiv1209.4503S} from the neutron star-white dwarf binary PSR J1738+0333 \cite{2012MNRAS.423.3316A} yielding $\left|\hat{\alpha}_1\right| = -0.4^{+ 3.7}_{-3.1}\times 10^{-5}$.
\subsubsection{$\alpha_2$}
Concerning $\alpha_2$, Nordtvedt \cite{1987ApJ...320..871N} inferred an upper bound of the order of $10^{-7}$ from the close alignment of the Sun's angular momentum with the total angular momentum of the Solar System. Nonetheless, as remarked by Shao and Wex \cite{2012arXiv1209.4503S}, such a result strongly depends on the assumption that the two angular momenta were aligned just after the formation of the Solar System about $5$ Gyr ago. If compared to the  well known planetary orbital motions and the dynamical forces determining them, the hypothesis on which the analysis by Nordtvedt \cite{1987ApJ...320..871N}  is based is relatively more speculative and less testable. Another potential issue of the Nordtvedt reasoning \cite{1987ApJ...320..871N}  may be the following one.

The orientation of the  Solar System's invariable plane \cite{2012A&A...543A.133S} is fixed by
the values of the celestial coordinates of its north pole: the right ascension $\alpha$ (RA) and the declination $\delta$ (DEC). At epoch J2000.0, they are \cite{2007CeMDA..98..155S}
\eqi\alpha_{\rm inv}=273.85^{\circ},\delta_{\rm inv}=66.99^{\circ};\eqf
thus, its normal unit vector $\bds{\hat{L}}$ is
\begin{align}
\hat{L}_x & = \cos\delta_{\rm inv}\cos\alpha_{\rm inv} = 0.03, \\ \nonumber \\
\hat{L}_y & = \cos\delta_{\rm inv}\sin\alpha_{\rm inv} = -0.39, \\ \nonumber \\
\hat{L}_z & = \sin\delta_{\rm inv} = 0.92.
\end{align}
The north pole of rotation of the Sun at J2000.0 is  characterized by
\cite{2007CeMDA..98..155S}
\eqi\alpha_{\odot}=286.13^{\circ},\delta_{\odot}=63.87^{\circ},\eqf
so that the Sun's spin axis $\bds{\hat{S}}^{\odot}$ is
\begin{align}
\hat{S}_x^{\odot} \lb{spinx} & = 0.122, \\ \nonumber \\
\hat{S}_y^{\odot} \lb{spiny} & = -0.423, \\ \nonumber \\
\hat{S}_z^{\odot} \lb{spinz} & = 0.897.
\end{align}
Thus, the angle  between the Sun's spin axis $\bds{\hat{S}}$ and the Solar System's total angular momentum $\bds{L}$ is
\eqi\theta_{LS}=5.97^{\circ}.\lb{angolo}\eqf
While the orientation of the invariable plane is nowadays quite accurate, being at  $\approx 0.1-0.3\ {\rm mas\ cty^{-1}}$ level \cite{2012A&A...543A.133S}, it is not so for the Sun's spin axis. Indeed, its determination is usually made in terms of the Carrington elements
\cite{Carr1863,AA013}
$i$, which is the angle from the Sun's equator to the ecliptic, and the longitude of the node $\mathrm{\Omega}$ of the Sun's equator with respect to the Vernal equinox $\curlyvee$ along the ecliptic, by means of
\begin{align}
\hat{s}^{\odot}_x \lb{Sx} & = \sin i\sin{\mathrm{\Omega}}, \\ \nonumber \\
\hat{s}^{\odot}_y \lb{Sy} & = -\sin i\cos{\mathrm{\Omega}}, \\ \nonumber \\
\hat{s}^{\odot}_z \lb{Sz} & = \cos i.
\end{align}
Beck and Giles \cite{2005ApJ...621L.153B} recently measured them
from time-distance helioseismology analysis of Dopplergrams from the Michelson Doppler Imager (MDI) instrument on
board the SOlar and Heliospheric Observatory (SOHO) spacecraft; the data span was $\Delta t^{'} = 5$ yr, from May 1996  through
July 2001. The resulting values are\footnote{After rotating \rfr{Sx}-\rfr{Sz}, calculated with \rfr{iCarr}-\rfr{OCarr}, from the ecliptic to the Earth's equator, \rfr{spinx}-\rfr{spinz} are obtained.}
\begin{align}
i \lb{iCarr}& = 7.155^{\circ}\pm 0.002^{\circ}, \\ \nonumber \\
{\mathrm{\Omega}} \lb{OCarr}& = 73.5^{\circ}\pm 1^{\circ}.
\end{align}
Maybe a better accuracy could be reached in a near future with a dynamical measurement from planetary orbital motions by exploiting the general relativistic Lense-Thirring effect \cite{2011arXiv1112.4168I}.
The figures of \rfr{iCarr}-\rfr{OCarr} imply an ability to observationally constrain putative rates of change of $\theta_{LS}\ton{i,{\mathrm{\Omega}},\alpha_{\rm inv},\delta_{\rm inv}}$ with a necessarily limited accuracy; it can approximately be evaluated as
\eqi\sigma_{\dot\theta_{LS}}\approx \rp{\sqrt{\ton{\derp{\theta_{LS}}{i}}^2\sigma^2_{i} + \ton{\derp{\theta_{LS}}{{\mathrm{\Omega}}}}^2\sigma^2_{{\mathrm{\Omega}}}}}{\Delta t^{'}}= 7.73\times 10^2\ {\rm arcsec\ yr^{-1}}.\eqf It is an important limiting factor when hypothesized secular rates $\dot\theta_{LS}$ are used to constrain parameters entering the models proposed to explain $\dot\theta_{LS}$.

It is just the case as far as $\alpha_2$ is concerned. Indeed, Nordtvedt \cite{1987ApJ...320..871N} noticed that one of the dynamical consequences of the first term in \rfr{hamtilde} is a torque $\mathbf{\tau}_{\alpha_2}$ causing a precession of the Sun's spin axis about $\bds w$ at a rate $\Xi_{\alpha_2}$ proportional to $\alpha_2$. Competing classical torques due to the planets of the Solar System induce an extremely low overall precession of the Sun's spin axis characterized by a rate \cite{1987ApJ...320..871N}
$\Xi_{\rm class}\approx 10^{-10}\ {\rm yr^{-1}}$. Under certain simplifying assumptions, Nordtvedt \cite{1987ApJ...320..871N} obtained
\eqi\sin\ton{\rp{\theta_{LS}}{2}}=\ton{\rp{\Xi_{\alpha_2}}{\Xi_{\rm class}}}\sin\ton{\rp{\Xi_{\rm class} t}{2}}.\lb{torquo}\eqf By assuming that the Sun's equator and the invariable plane were aligned at the birth of the Solar System 5 Gyr ago, and that the present-day value of \rfr{angolo} is due to the steady action of the aforementioned torques throughout the life of the Solar System in such a way that $\Xi_{\rm class} \Delta T/2\ll 1, \Delta T=5\ {\rm Gyr}$, \rfr{torquo} reduces to
\eqi \theta_{LS} \approx \Xi_{\alpha_2}\Delta T.\lb{limite}\eqf In other words, the PF $\alpha_2 -$induced spin precession would cause a secular rate of change $\dot\theta_{LS}$ whose magnitude is approximately equal to $\Xi_{\alpha_2}$. By posing $\Delta T=5\ {\rm Gyr}$, Nordtvedt \cite{1987ApJ...320..871N} inferred an upper limit of $\alpha_2$
as little as $\approx 10^{-7}$ from a hypothesized precession
\eqi\dot\theta_{LS}=\rp{5.97^{\circ}}{5\ {\rm Gyr}}=4\times 10^{-6}\ {\rm arcsec\ yr^{-1}}.\lb{pazzo}\eqf In fact, apart from the more or less speculative assumptions about a dynamical evolution spanning 5 Gyr, the observational uncertainty in actually measuring such a kind of time derivative of $\theta_{LS}$ should be  considered in deriving realistic constraints on $\alpha_2$. As we have seen, the errors in determining the Carrington elements of the Sun's spin axis yield an uncertainty in observationally constraining $\dot\theta_{LS}$ which may be about $8$ orders of magnitude larger than \rfr{pazzo}.

For  previous tests involving also other phenomena such as, e.g., Earth tides, see \cite{1976ApJ...208..881W,1993tegp.book.....W}.

Moving to the strong-field regime,  Shao and Wex \cite{2012arXiv1209.4503S} reported $\left|\hat{\alpha}_2\right|\leq 1.8\times 10^{-4}$ from the rate of change of the projected semimajor axis of the two pulsar-white dwarf binaries PSR J1012+5307 \cite{2001MNRAS.326..274L} and PSR J1738+0333 \cite{2012MNRAS.423.3316A}.
Earlier limits ($-0.3 < \hat{\alpha}_2 < 0.2$) were inferred by Wex and Kramer \cite{2007MNRAS.380..455W} by using the double pulsar PSR J0737-3039A/B \cite{doppia} made of two neutron stars. As remarked by Shao and Wex \cite{2012arXiv1209.4503S} themselves, a potential drawback of their latest constraint on $\hat{\alpha}_2$ is that it was obtained by combining data from two systems whose neutron stars have different masses. Moreover, a straightforward comparison with the previous test by  Wex and Kramer \cite{2007MNRAS.380..455W} is, in principle, problematic because the latter probed the interaction between two strongly self-gravitating objects, while the systems used by  Shao and Wex \cite{2012arXiv1209.4503S} contain white dwarfs.
\subsection{Bounds from recent Solar System planetary data}
\begin{table*}[ht!]
\caption{Supplementary precessions $\Delta\dot{\mathit{\Omega}}, \Delta\dot \varpi$ of
the longitudes of the node  and of the perihelion for some planets of the Solar System 
 estimated by Fienga et al. \cite{2011CeMDA.111..363F} with the INPOP10a ephemerides. \textcolor{black}{Actually, more recent versions of the INPOP ephemerides, named INPOP10e \cite{2013arXiv1301.1510F} and INPOP13a \cite{2013arXiv1306.5569V}, have been recently produced; no supplementary orbital precessions have yet been released for them.}
 The reference $\{x,y\}$ plane is the mean Earth's equator at J$2000.0$. The units are  milliarcseconds per century (mas cty$^{-1}$). \textcolor{black}{Fienga et al. \cite{2011CeMDA.111..363F} did not model PFE by keeping the values of the PPN parameters $\beta,\gamma$ fixed to their general relativistic values.
 Thus, their supplementary precessions $\Delta\dot{\mathit{\Omega}}, \Delta\dot \varpi$ effectively account for any kind of departures from standard dynamics in a  model-independent way. }
}\label{tavola}
\centering
\bigskip
\begin{tabular}{lll}
\hline\noalign{\smallskip}
&   $\Delta\dot \Om$ (mas cty$^{-1}$) & $\Delta\dot \varpi $ (mas cty$^{-1}$)  \\
\noalign{\smallskip}\hline\noalign{\smallskip}
Mercury & $1.4 \pm 1.8$ & $0.4 \pm 0.6$ \\
Venus & $0.2 \pm 1.5$ & $ 0.2\pm 1.5$ \\
Earth & $0.0\pm 0.9$ & $-0.2\pm 0.9$ \\
Mars & $-0.05\pm 0.13$ & $-0.04\pm 0.15$ \\
%
%
Saturn & $-0.1\pm 0.4$ & $0.15\pm 0.65$ \\
\noalign{\smallskip}\hline\noalign{\smallskip}
\end{tabular}
\end{table*}
According to Table \ref{tavola}, the present-day accuracy in constraining possible extra-precessions of the nodes and perihelia with respect to standard Newtonian/Einsteinian effects has reached the $\approx 1-0.5$ milliarcseconds per century (mas cty$^{-1}$) level for the inner planets of the Solar System and for Saturn.
This fact, in conjunction with the theoretical predictions of Appendix \ref{La1} and Appendix \ref{La2}, allows to infer tight constraints on both $\alpha_1$ and $\alpha_2$.

We note that Fienga et al. \cite{2011CeMDA.111..363F} neither explicitly modelled PFE nor simultaneously estimated $\alpha_1,\alpha_2$ along with the other solved-for parameters of their analysis. Thus, it may be speculated that the bounds obtained here by straightforwardly comparing our analytical predictions with the extra-precessions of Table \ref{tavola} are too optimistic since, e.g., a hypothetically existing PFE signal may have been partially removed from the residuals, being somewhat absorbed into the values of the estimated parameters such as, say, the planetary state vectors. In other words, there may be still room for larger values of $\alpha_1,\alpha_2$ than those inferred by us. Actually, we are not convinced that such an objection can invalidate our findings, although fitting  suitably modified dynamical models to the same planetary observations and  estimating $\alpha_1,\alpha_2$ \textcolor{black}{simultaneously with, e.g., the other PPN parameters} would certainly be a useful complementary approach \textcolor{black}{which could also broaden the range of applicability of the resulting constraints (see below)}. On the one hand, one could always argue that, even by explicitly solving for PFE, the resulting constraints on $\alpha_1,\alpha_2$ may still be impacted by any other sort of unmodelled/mismodelled forces, both standard and exotic. Indeed, a selection of the dynamical effects to be modelled and of the parameters entering them which can be practically estimated is always made in real data reduction. Thus, the effect of any sort of \virg{Russell teapots} may well creep into the the solved-for values of $\alpha_1,\alpha_2$ estimated in a full covariance analysis. On the other hand, the existing literature shows that, actually, our approach has been often adopted  in other studies using planetary data to constrain non-standard effects (Dark Matter \cite{2006IJMPD..15..615K,Pitjevi013}, MOND \cite{2011MNRAS.412.2530B}, general departures from inverse-square law \cite{2006MNRAS.371..626S}, etc.).  Nordtvedt himself \cite{1987ApJ...320..871N} based his analysis on already existing observations previously reduced by the astronomers who did not model PFE at all and  assumed a-priori the validity of general relativity; his constraint on $\alpha_2$ \cite{1987ApJ...320..871N} does not come from a fit of dynamical models including PFE. Moreover, we believe that the validity of our approach was indirectly confirmed, at least for certain hypothetical extra-forces, in independent analyses by some astronomers who explicitly dealt with certain anomalous effects of interest. Indeed, in the case of, e.g., the Pioneer anomaly \cite{1998PhRvL..81.2858A,2002PhRvD..65h2004A}, we concluded \cite{2008mgm..conf.2558I} that it could not be caused by  an exotic gravitational mechanism just by comparing the predicted planetary perihelion precessions due to a uniform acceleration radially directed towards the Sun with the limits of the anomalous planetary perihelion precessions obtained by some astronomers without explicitly modeling such a putative acceleration.  In subsequent researches, either ad-hoc modified dynamical planetary theories were fitted by some astronomers to data records of increasing length and quality getting quite negative results for values of the anomalous radial acceleration as large as the Pioneer one \cite{2008AIPC..977..254S,2009sf2a.conf..105F,2010IAUS..261..179S,2012sf2a.conf...25F}, or they explicitly modeled and solved for a constant, radial acceleration inferring admissible upper bounds \cite{2009IAU...261.0601F} not weaker than those obtained by us \cite{2011JCAP...05..019I}. On the contrary, to our knowledge, there are no published papers in which real planetary data were processed to constrain some exotic accelerations by explicitly estimating them which show a substantial discrepancy with the results obtained with our approach. A partial exception may be represented by the interesting work by Hees et al. \cite{2012CQGra..29w5027H} in which certain modified models of gravity were fitted to simulated observations and some dedicated parameters were determined. Nonetheless, its validity could well be limited just to the specific effect investigated and to the data simulation procedure adopted. Last but not least, it would be difficult to understand how  Le Verrier \cite{leverrier59} successfully measured the anomalous perihelion precession of Mercury by processing its observations with purely Newtonian models. The completely unmodelled general relativistic signature was not removed from the residuals, thus allowing Einstein \cite{1915SPAW...47..831E} to solve the puzzle of the  Mercury's anomalous motion just by comparing his theoretical precession with the existing observational determination. Finally, we remark that our constraints on $\alpha_1,\alpha_2$ should not be considered as too optimistic also because we assumed that the whole range of variation of the extra-precessions in Table \ref{tavola} is entirely due to PFE. \textcolor{black}{On the other hand, it must also be admitted that such a choice somewhat restricts the range of applicability of our results. Indeed, by using observationally-determined quantities such as the supplementary perihelion precessions of Table \ref{tavola} obtained by keeping the other PPN parameters fixed to their general relativistic values implies that the hypothetical PFE theory constrained by us deviates from general relativity in at most\footnote{\textcolor{black}{From this point of view, some distinctive symbols such as, say, a tilde may have been adopted for the PFE parameters actually constrained in this paper in much the same way as a hat was put on their strong-field values. We did not so to avoid a too cumbersome notation}.} $\alpha_1,\alpha_2$. }

As far as $\bds w$ is concerned, a natural choice for the preferred frame, common to the literature on preferred-frame effects \cite{1976ApJ...208..881W,1984grg..conf..365H,1987ApJ...320..871N,1992PhRvD..46.4128D,1994PhRvD..49.1693D,2012arXiv1209.4503S}, is the Cosmic Microwave Background (CMB). With such a choice one is quite plausibly assuming that the preferred frame is determined by the global matter distribution of the Universe, with the extra-components of the gravitational interaction ranging over scales at least comparable to the Hubble radius. Latest results from the Wilkinson Microwave Anisotropy Probe (WMAP) yield a peculiar velocity of the Solar System Barycenter (SSB) of \cite{2009ApJS..180..225H}
\eqi w = 369.0\pm 0.9\ {\rm km\ s^{-1}}, l = 263.99^{\circ}\pm 0.14^{\circ},b = 48.26^{\circ}\pm 0.03^{\circ},\lb{wmap}\eqf
where $l$ and $b$ are the Galactic longitude and latitude, respectively. Thus, in Celestial coordinates, the components \textcolor{black}{of the unit vector} $\bds{\hat{w}}$ are
\begin{align}
\wx \lb{wmapx} & = -0.970, \\ \nonumber \\
\wy & = 0.207, \\ \nonumber \\
\wz \lb{wmapz}& = -0.120.
\end{align}

By comparing the figures in Table \ref{tavola} with the theoretical predictions of Appendix \ref{La1} and Appendix \ref{La2}, computed with \rfr{wmapx}-\rfr{wmapz}, it is possible to infer
upper bounds on $\alpha_1$ and $\alpha_2$.
\begin{table*}[ht!]
\caption{Constraints on $\alpha_1$ and $\alpha_2$ obtained from a straightforward comparison of the figures of Table \ref{tavola} for the supplementary rates of the planetary  perihelia with the theoretical predictions of Appendix \ref{La1} and Appendix \ref{La2}, calculated with \rfr{wmap}. The tightest bounds comes from the perihelion of Earth ($\alpha_1$) and of Mercury ($\alpha_2$).
}\label{tavola2}
\centering
\bigskip
\begin{tabular}{lll}
\hline\noalign{\smallskip}
&  $\alpha_1$  & $\alpha_2$  \\
\noalign{\smallskip}\hline\noalign{\smallskip}
Mercury  & $(-3\pm 5)\times 10^{-6}$ & $(4\pm 6)\times 10^{-6}$ \\
Venus  & $(-0.1\pm 1.1)\times 10^{-5}$ & $(-0.7\pm 5.7)\times 10^{-5}$ \\
Earth  & $(0.8\pm 4)\times 10^{-6}$ & $(-0.8\pm 3.7)\times 10^{-5}$ \\
Mars  & $(-0.8\pm 2.9)\times 10^{-5}$  & $(0.4\pm 1.5)\times 10^{-5}$ \\
Saturn &  $(-1.94\pm 8.41)\times 10^{-4}$ &  $(1.95\pm 8.5)\times 10^{-4}$ \\
\noalign{\smallskip}\hline\noalign{\smallskip}
\end{tabular}
\end{table*}
It turns out that, in general, the nodes yield  weaker constraints than the perihelia, especially as far as $\alpha_1$ is concerned.
Thus, in Table \ref{tavola2} only the bounds from the perihelia are displayed.
The most stringent constraint on $\alpha_1$ comes from the perihelion of Earth, amounting to $\left|\alpha_1\right|\lessapprox 10^{-6}$. The perihelion of Mercury yields the tightest constraint for $\alpha_2$, which is of the order of $\left|\alpha_2\right|\lessapprox 10^{-6}$ as well.
In principle, the bounds of Table \ref{tavola2} may be biased by certain standard dynamical effects which were not modeled at all by Fienga et al. \cite{2011CeMDA.111..363F}, such as the general relativistic Lense-Thirring effect \cite{LenseThirring1918}, thus impacting the supplementary rates $\Delta\dot\varpi$ of Table \ref{tavola}. The gravitomagnetic precession should not be neglected since its expected value ($-2.0$ mas cty$^{-1}$) is larger than the current uncertainty in the Mercury's perihelion extra-rate ($0.6$ mas cty$^{-1}$); see \cite{2011arXiv1112.4168I} for a recent discussion of this aspect. Another standard effect which may a-priori impact the bounds in Table \ref{tavola2} through its lurking in Table \ref{tavola}  is the Newtonian perihelion precession due to the Sun's quadrupole moment $J_2$. Although its dynamical effect was fully modeled by Fienga et al. \cite{2011CeMDA.111..363F}, a lingering uncertainty of \cite{2011EPJH...36..407R} $\approx 10\%$ still affects $J_2$ in such a way that a resulting mismodelled perihelion precession certainly contributes to the figures in Table \ref{tavola}. The availability of more than one supplementary perihelion rate is a great advantage since it allows to set up a suitable linear combination involving the perihelia of the four inner planets able to separate, by construction, $\alpha_1$, $\alpha_2$ from $J_2$ and the Lense-Thirring effect. This method has been used in a number of papers in the literature; it was proposed for the first time in
\cite{2005A&A...433..385I,2005A&A...431..385I}, as far as the Solar System planetary scenario is concerned. From the following linear system of four equations in the four unknowns $\alpha_1,\alpha_2,J_2,\mu_{\rm LT}$
\eqi\Delta\dot\varpi^{j} = \alpha_1\dot\varpi^{j}_{.\alpha_1} + \alpha_2\dot\varpi^{j}_{.\alpha_2} + J_2\dot\varpi^{j}_{. J_2} + \mu_{\rm LT}\dot\varpi^{j}_{.\rm LT},\ j={\rm Mercury\ldots Mars},\lb{sistemino}\eqf
where the coefficients $\dot\varpi_{.}$ are the analytical expressions of the pericenter precessions caused by the effects considered\footnote{General expressions for the $J_2$ and Lense-Thirring pericenter precessions can be found in \cite{2011PhRvD..84l4001I}. The coefficient $\mu_{\rm LT}$ is one in general relativity.}, it is possible to obtain
\begin{align}
\alpha_1 \lb{upa1} & = (-1\pm 6)\times 10^{-6}, \\ \nonumber \\
\alpha_2 \lb{upa2} & = (-0.9\pm 3.5)\times 10^{-5}, \\ \nonumber \\
\textcolor{black}{J_2} \lb{j2cazzo} & \textcolor{black}{= \left(1.4 \pm 4.1\right)\times 10^{-7}}, \\ \nonumber \\
\textcolor{black}{\mu_{\rm LT}} \lb{LTcazzo}& \textcolor{black}{= 8\pm 24}.
\end{align}
By construction, \textcolor{black}{the figures of \rfr{upa1}-\rfr{upa2}} are not a-priori biased by any mismodeling in $J_2$ and by the Lense-Thirring effect, independently of their values. \textcolor{black}{Incidentally, let us notice that, although not specifically designed for determining/constraining $J_2, \mu_{\rm LT}$, the system of \rfr{sistemino} yields figures for them (\rfr{j2cazzo} and \rfr{LTcazzo}) which are compatible with their expected values.} 

The constraints of Table \ref{tavola2} and of \rfr{upa1}-\rfr{upa2} will likely be further improved in the next few years. Indeed, our knowledge of the orbit of Mercury will be greatly enhanced when the entire data set of the MESSENGER spacecraft \cite{2007SSRv..131....3S}, which was inserted in orbit around Mercury in March 2011 for a year-long science phase extended until March 2013, will be processed. Another planned mission to Mercury is BepiColombo \cite{2010P&SS...58....2B}: it will be launched in 2015 and should reach its target in 2022. One of its experiment is MORE \cite{2009AcAau..65..666I}; it should further improve the determination of the orbit of Mercury \cite{2007IJMPD..16.2117I}. \textcolor{black}{For existing studies concerning, among other things, the opportunity of constraining also $\alpha_1$ and $\alpha_2$ with BepiColombo using different approaches, see \cite{2002PhRvD..66h2001M,2007PhRvD..75b2001A}. The  covariance analysis in \cite{2002PhRvD..66h2001M}, based on a full cycle simulation of the BepiColombo radio science experiments, shows that $\alpha_1$ and $\alpha_2$ can be constrained within\footnote{\textcolor{black}{They are realistic, non-formal uncertainties \cite{2002PhRvD..66h2001M}.}} $\simeq 8\times 10^{-6}$ and $\simeq 10^{-6}$, respectively. The \virg{modified worst-case} error analysis in \cite{2007PhRvD..75b2001A}, which is a quite different approach with respect to that adopted in \cite{2002PhRvD..66h2001M}, predicts accuracies of the order of $\simeq 2.1\times 10^{-5}-8.6\times 10^{-7}\ (\alpha_1)$ and $2.9\times 10^{-6}-1.2\times 10^{-6}\ (\alpha_2)$ depending on the assumed mission duration ranging from 1 yr to 8 yr. Such constraints are not too far from those in Table \ref{tavola2} and in \rfr{upa1}-\rfr{upa2}, especially as far as $\alpha_1$ is concerned. }
\section{Summary and conclusions}\lb{concludi}
The long-term, i.e. averaged over one orbital revolution, precessions of the orbital elements of the relative motion of the components of a binary system due to non-zero values of the PPN parameters $\alpha_1$ and $\alpha_2$ inducing preferred-frame effects were analytically worked out to first-order with the perturbative Lagrange planetary equations. We did not recur to a-priori simplifying assumptions on the  preferred-frame velocity $\bds w$ and on the binary's orbital geometry. The analytical expressions of the precessions were successfully tested with a numerical integration which yielded the same results. It was also numerically checked that, to the order of approximation used, the contact orbital elements used in the analytical calculation are actually tangent, i.e. osculating, to the perturbed trajectory.

We critically discussed the existing bounds on $\alpha_1,\alpha_2$. We remarked that the $\mathcal{O}\ton{10^{-7}}$ constraint on $\alpha_2$ inferred by Nordtvedt from a hypothesized secular precession of the angle between the Sun's equator and the invariable plane of the Solar System  throughout its entire existence should be considered as optimistic. Indeed, it did not take into account the measurement limits in actually constraining such a precession from observations in view of the lingering uncertainties in the Carrington elements $i$ and, especially, $\mathrm{\Omega}$ which determine the orientation of the Sun's spin axis.

\textcolor{black}{Deviating from the approach followed so far in literature,} we used the supplementary node and perihelion secular precessions $\Delta\dot{\mathit{\Omega}},\Delta\dot\varpi$ of some planets of the Solar System, recently determined by some astronomers from an analysis of a centennial data record in building the INPOP10a ephemerides, to preliminarily constrain both $\alpha_1$ and $\alpha_2$. Indeed, such supplementary precessions, \textcolor{black}{by construction,} account for any unmodelled/mismodelled dynamical effects with respect to the standard Einstein-Newton gravity \textcolor{black}{in the sense that the  PPN parameters of the post-Newtonian forces modelled were kept fixed to their standard general relativistic values}. After having noticed that the perihelia give tighter constraints than the nodes,
we linearly combined the supplementary perihelion precessions $\Delta\dot\varpi$ of Mercury, Venus, Earth, Mars to simultaneously solve for $\alpha_1,\alpha_2$, and for the Sun's Lense-Thirring effect and oblateness $J_2$ as well. Indeed, they were unmodelled/mismodelled in the INPOP10a-based analysis and represent competing effects for the $\alpha_1,\alpha_2-$induced precessions we are interested in.
We obtained  $\alpha_1 = (-1\pm 6)\times 10^{-6},\alpha_2 = (-0.9\pm 3.5)\times 10^{-5}$. Our results, which retain a general validity for all the weak-field scenarios \textcolor{black}{departing from general relativity just as far as PFE are concerned}, are based on well tested dynamical effects pertaining known Solar System's objects, and do not rely upon speculative assumptions about hypothesized phenomena for which no data actually exist.  \textcolor{black}{On the other hand, we inferred them by assuming that the ranges $\Delta\dot\varpi$ are entirely due to $\alpha_1,\alpha_2$ themselves; such a choice may somewhat limit the range of applicability of our results to just those PFE theories deviating from general relativity in at most $\alpha_1,\alpha_2$ themselves}. As a complementary approach which, in principle, could be followed in dedicated full covariance analyses, the preferred-frame effects should be explicitly included in purposely modified dynamical models which should be fitted to the planetary data set, and $\alpha_1,\alpha_2$ should be simultaneously estimated along with \textcolor{black}{the} other \textcolor{black}{PPN} parameters. Nonetheless, its practical implementation is not at all trivial. \textcolor{black}{It is worthwhile noticing that we did not limit ourselves just to the node and the pericenter, having computed the long-term rates of change also for the other Keplerian orbital elements exhibiting peculiar signatures with respect to other potentially competing dynamical effects. They could turn out useful in future if and when the astronomers will determine the ranges for their supplementary rates as well. Analogous considerations hold also for other astronomical systems.}

Future improvements in the orbit determination of Mercury are expected in the near and mid future from MESSENGER and BepiColombo missions; they should allow to get tighter constraints on $\alpha_1,\alpha_2$. \textcolor{black}{In this respect, it is remarkable that the constraints expected from BepiColombo for both $\alpha_1$ and $\alpha_2$ in independent analyses are close to those inferred by us in this paper, especially as far as $\alpha_1$ is concerned.}

\renewcommand\appendix{\par
\setcounter{section}{0}%
\setcounter{subsection}{0}%
\setcounter{table}{0}
\setcounter{figure}{0}
\setcounter{equation}{0}
\gdef\thetable{\Alph{table}}
\gdef\thefigure{\Alph{figure}}
\gdef\theequation{\Alph{section}.\arabic{equation}}
\section*{Appendices}
\gdef\thesection{\Alph{section}}
\setcounter{section}{0}}

\appendix
\section{The $\alpha_1$ precessions}\lb{La1}
The orbital precessions induced by \rfr{ham1} can be conveniently computed by considering separately each of the three terms entering \rfr{ham1}.
The following analytical expressions are exact in the sense that no a priori assumptions on the spatial orientation of $\bds w$ were assumed. Moreover, the limits of small eccentricity and inclination were not assumed as well. \textcolor{black}{It is worthwhile noticing that the mean anomaly at the epoch $\mathcal{M}_0$ does not enter the following averages of the pertubing Hamiltonians. Thus, according to \rfr{pippa1}-\rfr{pippa2}, the semimajor axis $a$ remains unaffected, while the eccentricity $e$ may change only if the averaged Hamiltonians depend on the argument of pericenter $\omega$.}
\subsection{The $v^2$ term}\lb{La1_v2}
\textcolor{black}{The average of the term proportional to the square of the two-body relative velocity in \rfr{ham1} is
\eqi
\ang{{\mathcal{H}}_{\alpha_1}^{v^2}} = \rp{\alpha_1 G \nk^2 m_{\rm A}m_{\rm B} a}{2c^2 M}\ton{1 - \rp{2}{\sqrt{1-e^2}}}.
\eqf
Thus,
the long-term precessions caused by it are all zero, apart from the longitude of the pericenter.
Its} rate of change
is
\begin{align}
\ang{\dert\varpi t} \lb{dvarpidt1v2} & = \rp{\alpha_1 G m_{\rm A}m_{\rm B}\nk}{c^2 a\ton{1 - \ee}M}.
%
\end{align}
\subsection{The $w^2$ term}\lb{La1_w2}
All the precessions due to the term proportional to $w^2$ in \rfr{ham1} vanish. \textcolor{black}{Indeed, its average is
\eqi\ang{{\mathcal{H}}_{\alpha_1}^{w^2}} = \rp{\alpha_1 G w^2 M }{2a c^2}.\eqf
}
\subsection{The $\ton{\bds v\bds\cdot\bds w}$ term}\lb{La1_Dm}
\textcolor{black}{The average of the mixed term of \rfr{ham1}, proportional to $\Delta m$,
is
\begin{align}
\ang{{\mathcal{H}}_{\alpha_1}^{\Delta m}} \nonumber \lb{uffaa} & = \rp{\alpha_1 w G\nk\Delta m\ton{-1 + e^2 + \sqrt{1-e^2}}}{2c^2e\sqrt{1 - e^2}}\qua{\wz\sI\co + \right. \\ \nonumber \\
 & + \left.\cI\co\ton{\wy\cO -\wx\sO} -\so\ton{\wx\cO + \wy\sO} }.
\end{align}
It} yields non-vanishing long-term variations for all the orbital elements, apart from $a$.
They are
\begin{align}
\ang{\dert e t} \lb{dedt1Dm} \nonumber & =  -\rp{\alpha_1 w G\Delta m \kle }{2 c^2 a^2 e^2}\grf{\co\wpl + \right. \\ \nonumber \\
& + \left.\so\qua{\wz\sI + \cI\wmn}}, \\ \nonumber \\
\ang{\dert{I} t}  \lb{dIdt1Dm} \nonumber & =  \rp{\alpha_1 w G \Delta m \kle \so}{2 c^2 a^2 e\ton{1 - \ee}}\qua{\wz\cI -\right. \\ \nonumber \\
& - \left. \sI\wmn}, \\ \nonumber \\
\ang{\dert{\mathit{\Omega}} t}  \lb{dOdt1Dm} \nonumber & =  -\rp{\alpha_1 w G\Delta m\kle \csc I\co}{2 c^2 a^2 e\ton{1 - \ee}}\qua{\wz\cI -\right. \\ \nonumber \\
& - \left. \sI\wmn}, \\ \nonumber \\
\ang{\dert{\varpi} t}  \lb{dvarpidt1Dm} \nonumber & =  \rp{\alpha_1 w G\Delta m }{2 c^2 a^2 e}\grf{ \ton{\rp{-1 + \sq}{\ee}}\qua{\wz\sI\co + \right.\right. \\ \nonumber \\
\nonumber & + \left.\left. \cI\co\wmn - \right.\right. \\ \nonumber \\
\nonumber & - \left.\left. \so\wpl } - \right. \\ \nonumber \\
\nonumber & - \left. \ton{\rp{-1 + \ee + \sq}{1 - \ee}}\tan\ton{\rp{I}{2}}\co\qua{\wz\cI - \right.\right. \\ \nonumber \\
& - \left.\left. \sI\wmn}  }.
%
\end{align}

It turns out that the leading term in \rfr{dedt1Dm} is of order $\mathcal{O}\ton{e^0}$ in the eccentricity, while the next-to-leading order one is of order $\mathcal{O}\ton{e^2}$. As far as $I$ and $\mathit{\Omega}$ are concerned, their leading order terms are of order $\mathcal{O}\ton{e}$, while the first non-vanishing terms of higher order in $e$ are of order $\mathcal{O}\ton{e^3}$. The pericenter precession of \rfr{dvarpidt1Dm} has a term of order $\mathcal{O}\ton{e^{-1}}$, while the next ones are of order $\mathcal{O}\ton{e}$ and $\mathcal{O}\ton{e^3}$, respectively.
\section{The $\alpha_2$ precessions}\lb{La2}
The orbital precessions due to $\alpha_2$ are the same as of Appendix \ref{La1} rescaled by $-\alpha_2/\alpha_1$, and by those coming from \rfr{hamtilde}. Also in this case, analytical expressions exact in $\bds w,e,I$ are obtained.
\subsection{The $v_r^2$ term}\lb{La2_vr2}
\textcolor{black}{The average of the term proportional to $v_r^2$ in \rfr{hamtilde} is
\eqi\ang{{\widetilde{\mathcal{H}}}_{\alpha_2}^{v_r^2}} = \rp{\alpha_2 G\nk^2 m_{\rm A}m_{\rm B}}{2c^2 M}\ton{1 - \rp{1}{\sqrt{1-e^2}}}.\lb{skassa}\eqf
As a consequence, the long-term precessions induced by \rfr{skassa} vanish, apart from $\varpi$.}
Its rate
is
\begin{align}
\ang{\dert\varpi t} \lb{dvarpidt2vr2} & = \rp{\alpha_2 G m_{\rm A}m_{\rm B}\nk}{2c^2 a\ton{1 - \ee}M}.
%
\end{align}
\subsection{The $w_r^2$ term}\lb{La2_wr2}
\textcolor{black}{The average of the term proportional to  $w_r^2$ in \rfr{hamtilde} is
\begin{align}
\ang{{\widetilde{\mathcal{H}}}_{\alpha_2}^{w_r^2}} \nonumber & =\rp{\alpha_2 w^2 \nk^2 a^2}{2c^2 e^2}\grf{-\wx^2\sqrt{1-e^2} \cos ^2 I  \sin ^2\omega \sin ^2\Om + \right. \\ \nonumber \\
\nonumber & + \left. \wx^2\sqrt{1-e^2} \cos  I  \sin  2\omega  \sin  2\Om  - 2 \wx\wy\sqrt{1-e^2} \cos  I  \cos  2\Om  \sin  2\omega +\right.\\\nonumber \\
\nonumber & + \left. \wx\wz\sqrt{1-e^2} \sin  2I  \sin ^2\omega \sin \Om + \wx\wy\sqrt{1-e^2} \cos ^2 I  \sin ^2\omega \sin  2\Om   + \right.\\\nonumber \\
\nonumber & + \left.\wx\wy\sqrt{1-e^2}  \sin ^2\omega \sin  2\Om + \left[\wx^2\left(e^2+\sqrt{1-e^2}-1\right) -\right.\right.\\ \nonumber \\
\nonumber &-\left.\left. \wy^2\left(\sqrt{1-e^2}-1\right)  \cos ^2 I \right] \cos ^2\Om  \sin ^2\omega - \wz^2\sqrt{1-e^2} \sin ^2 I  \sin ^2\omega +\right.\\ \nonumber \\
\nonumber & + \left. \wy^2\sqrt{1-e^2}  \sin ^2\omega \sin ^2\Om -\wy \cos \Om  \sin ^2\omega\left[\wz\left(\sqrt{1-e^2}-1\right)  \sin  2I +\right.\right. \\ \nonumber \\
\nonumber & + \left.\left. \wx \left(-2 e^2+\cos  2I +3\right) \sin \Om \right]+\sin ^2\omega \left(\wz^2\sin ^2 I +\right.\right. \\ \nonumber \\
\nonumber & +\left.\left. \sin \Om  \left(\left(\wy^2\left(e^2 - 1\right) +\wx^2 \cos ^2 I \right) \sin \Om -\wx \wz \sin  2I \right)\right)+\right.\\ \nonumber \\
\nonumber &+ \left. \wy \cos  I  \sin  2\omega  \left(2 \wx \cos  2\Om +\wy \sin  2\Om \right)+\right.\\ \nonumber \\
\nonumber & + \left.\cos ^2\omega \left(-2 \wx \wz\cos  I  \sin  I  \sin \Om  e^2+\left(\wy^2\left(e^2+\sqrt{1-e^2}-1\right)  \cos ^2 I-\right.\right.\right.\\ \nonumber \\
\nonumber &- \left.\left.\left. \wx^2\left(\sqrt{1-e^2}-1\right) \right)\cos ^2\Om + \wz^2\left(e^2+\sqrt{1-e^2}-1\right)  \sin ^2 I -\right.\right.\\ \nonumber \\
\nonumber &-\left.\left. \wy^2\sqrt{1-e^2}  \sin ^2\Om + \wx^2\sqrt{1-e^2}  \cos ^2 I  \sin^2\Om +\right.\right. \\ \nonumber \\
\nonumber &+\left.\left. \left(\wy^2+\left(e^2-1\right) \wx^2 \cos ^2 I \right) \sin ^2\Om + \wx \wz\ton{1-\sqrt{1-e^2}}  \sin  2I  \sin \Om  +\right.\right.\\ \nonumber \\
\nonumber &+ \left.\left. \wy\wz \cos \Om  \left(\left(e^2+\sqrt{1-e^2}-1\right) \sin  2I -\right.\right.\right.\\ \nonumber \\
\nonumber &-\left.\left.\left. \wx \left(2 e^2\cos ^2 I +\sqrt{1-e^2} (\cos  2I +3)\right) \sin \Om \right)+\wx \wy \left(\cos ^2 I +1\right) \sin  2\Om \right)+\right.\\ \nonumber \\
\nonumber & +\left. 2 \sin 2\omega \left(\frac{1}{2} \cos  I  \left(\left(\wx^2\left(e^2-2\right) - \wy^2\left(e^2+2 \sqrt{1-e^2}\right) \right) \sin  2\Om -\right.\right.\right.\\ \nonumber \\
&-\left.\left.\left. 2e^2 \wx \wy \cos  2\Om \right) - \wz\left(e^2 + 2 \sqrt{1-e^2}-2\right)  \sin  I  (\wx \cos \Om +\wy \sin \Om )\right) }.
\end{align}
The resulting non-vanishing long-term rates of change of the orbital elements are as follows.
}

\begin{align}
\ang{\dert e t} \lb{dedt2wr2} \nonumber & = \rp{\alpha_2 w^2 n_{\rm b}\sqrt{1-\ee}\ton{- 2 + \ee + 2\sqrt{1-\ee}}}{8 c^2  e^3  }\cdot \\ \nonumber \\
\nonumber &\cdot  \grf{ -8 \wz \sI\coo\ton{\wx\cO + \wy\sO} +\right. \\ \nonumber \\
\nonumber & + \left. 4\cI\coo\qua{-2\wx \wy\cOO + \ton{\wx^2 - \wy^2}\sOO} +\right.\\ \nonumber \\
\nonumber & + \left.  \soo\qua{ \ton{\wx^2 - \wy^2}\ton{3 + \cII}\cOO + \right.\right. \\ \nonumber \\
\nonumber & + \left.\left. 2\sin^2 I\ton{\wx^2 + \wy^2 - 2 \wz^2} - 4\wz\sII\ton{\wy\cO - \wx\sO } + \right.\right. \\ \nonumber \\
 & + \left.\left. 2 \wx \wy \ton{3 + \cII}\sOO}}, \\ \nonumber \\
\ang{\dert I t} \lb{dIdt2wr2}\nonumber & =  \rp{\alpha_2 w^2 n_{\rm b}}{2 c^2  \ee \sq }\qua{\wz\cI - \sI\wmn}\cdot \\ \nonumber \\
\nonumber &\cdot \grf{- 2\sq\wx\cO\sin^2\omega + \right. \\ \nonumber  \\
\nonumber & + \left. 2\sq\wy\cI\cO\soo - \right.\\ \nonumber \\
\nonumber & - \left. 2\sq\wy\sO\sin^2\omega + \right. \\ \nonumber \\
\nonumber & + \left. 2\sq\cos^2\omega\wpl - \right. \\ \nonumber \\
\nonumber & - \left. 2\coo\wpl - \right. \\ \nonumber \\
\nonumber & - \left. 2\ee\sin^2\omega\wpl + \right. \\ \nonumber \\
\nonumber & + \left. 2\sq\soo\qua{\wz\sI -\wx\cI\sO} + \right. \\ \nonumber \\
& + \left. \ton{-2 + \ee} \soo\qua{\wz\sI + \cI\wmn}}, \\ \nonumber \\
\ang{\dert{\mathit{\Omega}} t} \lb{dOdt2wr2} \nonumber & =  -\rp{\alpha_2 w^2 n_{\rm b}\csc I}{2 c^2  \ee\sq  }\qua{\wz\cI - \sI\wmn}\cdot \\ \nonumber \\
\nonumber &\cdot \grf{\wz\sI\qua{\ee + \ple\coo} + \right. \\ \nonumber \\
\nonumber & + \left.\cI\qua{\ee + \ple\coo}\wmn - \right. \\ \nonumber \\
& - \left. \ple\soo\wpl}, \\ \nonumber \\
\ang{\dert \varpi t} \lb{dvarpidt2wr2} \nonumber & =  \rp{\alpha_2 w^2 n_{\rm b}}{8 c^2  e^4  }\cdot\\ \nonumber \\
\nonumber &\cdot \grf{\ple\cdot\right.\\ \nonumber  \\
\nonumber &\cdot  \left.\qua{\wq\cp\cOO\coo + \right.\right. \\ \nonumber \\
\nonumber & + \left.\left. 2\wl\sin^2 I\coo - \right.\right.\\ \nonumber \\
\nonumber & - \left.\left. 4\wz\sII\wmn\coo + \right.\right.\\ \nonumber \\
\nonumber & + \left.\left. 2\wx\wy\cp\sOO\coo + \right.\right.\\ \nonumber \\
\nonumber & + \left.\left. 8\wz\sI\wpl\soo + \right.\right.\\ \nonumber \\
\nonumber & + \left.\left. 8\wx\wy\cI\cOO\soo - \right.\right.\\ \nonumber \\
\nonumber & - \left.\left. 4\wq\cI\sOO\soo} - \right. \\ \nonumber \\
\nonumber & - \left. \rp{4\ee\tan\ton{\rp{I}{2}}}{\sq}\qua{\wz\cI - \sI\wmn}\cdot\right.\\ \nonumber \\
\nonumber &\cdot \left.\qua{\wz\ton{\ee + \ple\coo}\sI + \right.\right. \\ \nonumber \\
\nonumber & + \left.\left. \cI\ton{\ee + \ple\coo}\wmn - \right.\right.\\ \nonumber \\
& - \left.\left. \ple\wpl\soo  }}.
\end{align}

In the limit $e\rightarrow 0$, the leading terms in  \rfr{dedt2wr2}-\rfr{dvarpidt2wr2}  are of order $\mathcal{O}\ton{e^0}$. The next non-vanishing terms are of order $\mathcal{O}\ton{e^2}$ for all the elements, with the exception of $e$ whose next-to-leading order term is of order $\mathcal{O}\ton{e}$.
\subsection{The $v_r w_r$ term}\lb{La2_Dm}
The mixed term of \rfr{hamtilde}, proportional to $\Delta m$, causes non-vanishing long-term changes for all the orbital elements, apart from $a$.
They are formally identical to those in Appendix \ref{La1_Dm} with $\alpha_1$ replaced by $\alpha_2$ \textcolor{black}{since it turned out that $\ang{\widetilde{\mathcal{H}}_{\alpha_2}^{\Delta m}}$ is identical to $\ang{\mathcal{H}_{\alpha_1}^{\Delta m}}$ in \rfr{uffaa}.} Thus, $\Delta m$ does not contribute to the overall orbital perturbations due to $\alpha_2$ because the precessions coming from the $\Delta m$ term in \rfr{hamtilde} are canceled by those coming from the $\Delta m$ term in the first piece of \rfr{ham2}.

The total $\alpha_2-$induced precessions are made up of the $\ton{w/c}^2$ terms of Appendix \ref{La2_wr2}, proportional to $a^{-3/2}$, and of the terms containing the product of the masses which are completely negligible in the Solar System. In the case of a tight binary pulsar with, say, $m_{\rm A} \approx m_{\rm B}\approx M/2, a \approx 9.0\times 10^5\ {\rm km}$, the precessions of Appendix \ref{La1_v2}, rescaled by $-\alpha_2/\alpha_1$, and Appendix \ref{La2_vr2} are proportional to $\ton{v_{\rm orb}/c}^2$, with $v_{\rm orb}\approx 600\ {\rm km\ s^{-1}}$; for typical values of $w$ (see Section \ref{osserva}), they are dominant with respect to the precessions of Appendix \ref{La2_wr2}.

\bibliography{pfebib,QUMONDbib}{}

\end{document}